\documentclass{ewass_ss4proc}

\usepackage{kantlipsum}

\editors{Emeline Bolmont \& Sergi Blanco-Cuaresma}
\editorsurl{www.emelinebolmont.com | www.blancocuaresma.com/s/}
\publisher{Zenodo}
\conference{EWASS Special Session 4 (2017): Star-planet interactions}
\conferencedate{2017}

\title{Cosmic Rays near Proxima Centauri b}
\author{Alexei Struminsky,$^{1,2}$ 
        Andrei Sadovski,$^{1,3}$ and
        Anatoly Belov, $^{4}$}

\affiliation{$^{1}$ Space Research Institute, Moscow, Russia \\
			 $^{2}$ Moscow Institute of Physics and Technology, Dolgoprudny, Moscow Region, Russia \\
			 $^{3}$ National Research University Higher School of Economics, Moscow, Russia \\
			 $^{4}IZMIRAN, Troitsk, Moscow, Russia $}

\shorttitle{Cosmic Rays near Proxima b}
\shortauthors{Alexei Struminsky, Andrei Sadovski \& Anatoly Belov}

\abs{Cosmic rays are an important factor of space weather determining radiation conditions near the Earth and it seems to be essential to clarify radiation conditions near extrasolar planets too. Last year a terrestrial planet candidate was discovered in an orbit around Proxima Centauri. Here we present our estimates on parameters of stellar wind from the Parker model, possible fluxes and fluencies of galactic and stellar cosmic rays based on the available data of the Proxima Centauri activity and its 
magnetic field. We found that galactic cosmic rays will be practically absent near Proxima~b up to energies of 1~TeV due to the modulation by the stellar wind. Stellar cosmic rays may be accelerated in Proxima Centauri events, which are able to permanently maintain density of stellar cosmic rays in the astrosphere comparable to low energy cosmic ray density in the 
heliosphere. Maximal proton intensities in extreme Proxima events should be by 3--4 orders more than in solar events. }

\begin{document}

\maketitle

\section{Introduction}
Kepler discoveries of new extrasolar planets give great impetus for discussions of life conditions and possible conditions for habitable zone (see for ex. \citep{Anglada2016,Garraffo2016,Griessmeier2015, Griessmeier2016} and references within). Cosmic rays as a factor of space weather were considered only by one group, possibly, their first and most cited work in this regard is \citet{Griessmeier2005}.

The dependence of the Galactic cosmic rays (GCR) induced radiation dose on the strength of the planetary magnetic field and its atmospheric depth were considered in different papers \citep{Atri2013,Griessmeier2015}. \citet{Griessmeier2016} studied 
atmospheric implications of cosmic rays near extrasolar Earth-like planets. The authors of the cited works supposed that for such planets the GCR rays flux can be regarded as an isotropic and approximately constant, since at the Earth orbit only the flux of low energy particles is slightly modulated by the solar activity. Low modulation of GCR means also a weak dependence of GCR flux on the orbital distance. However since stellar wind velocity and 
magnetic field as well as an activity of other stars (especially dwarfs as in the case of Proxima~b) might be considerably higher in comparison with solar values, the modulation of GCR might be much stronger. \citet{Vidotto2014} mentioned the GCR modulation as a possible effect of the magnetized stellar wind, but without any estimate. \citet{Scherer2015} considered the modulation of GCR flux inside an astrosphere of $\lambda $~Cephei. This is an O star and its cold stellar wind has another nature than the hot wind of active dwarfs. According to Schrere et al. (2015) the modulation in astrospheres of O-B stars affects particles up to 100~TeV. 
Modulation of GCR by astrospheres of dwarfs is not considered yet.

A similar problem of the GCR modulation  near the archean Earth was considered by \citet{Scherer2002} and \citet{Cohen2012}. \citet{Scherer2002} demonstrated by quantitative modeling that a change of the interstellar medium surrounding the heliosphere triggers significant changes of planetary environments caused by enhanced fluxes of neutral atoms as well as by the increased cosmic ray fluxes. \citet{Cohen2012} showed that  the GCR flux near the Archean Earth (for the early Sun) would has greatly reduced than is the case today mainly due to the shorter solar rotation period and tighter winding of the Parker spiral, and to the different surface distribution of the more active solar magnetic field. 

Since stellar cosmic rays are not detectable (or distinguishable) far away from their parent star they are step-sons of cosmic ray physics, generally they are mentioned as a possible component of CR ones per ten years \citep{Unsold1957,Edwards1971,Lovell1974,Mullan1979,Kopysov2005,Stozhkov2011,Struminsky2017}. Stellar cosmic rays (SCR) were considered in many papers \citep{Tabataba2016,Atri2017,Struminsky2017} as an important factor of space weather in a habitable zone of star. Since the details of SCR spectrum is unknown to model the effect of SCR one may use spectra of well known solar events \citep{Atri2017} or average spectrum of solar proton events \citep{Tabataba2016}. Another approach is to base on general physical principles \citep{Struminsky2017} assuming solar-stellar analogies, which is not based on near Earth observations of solar cosmic rays.

The red dwarf Proxima Centauri ($\alpha$ Centauri~C, GL~551, HIP~70890 or simply Proxima) is the Sun's closest (1.295~parsecs) stellar neighbor and one of the best-studied low-mass stars. It has an effective temperature of only around 3,050~K, a luminosity of $0.15L_\odot$ (the index $\odot$ determines the Sun parameters), a measured radius of $0.14R_\odot$ and a mass of about $0.12M_\odot$ (GJ 551; dM5.5e) (see \citet{Anglada2016}). \citet{Anglada2016} reported the presence of a small planet with a minimum mass of about 1.3 Earth masses orbiting Proxima with a period of approximately 11.2~days at a semi-major-axis distance of around 0.049~AU so it should  rotate in the star's habitability zone. 

Proxima is considered as a moderately active star and its quiescent activity levels and X-ray luminosity~are comparable to those of the Sun. \citet{Wargelin2017} summarized several years of optical, UV, and X-ray observations of Proxima Centauri. They confirmed previous reports of an 83-day rotational period and find strong evidence for a 7-year stellar cycle, along with indications of differential rotation at about the solar level. The very long 
rotation period of Proxima Centauri renders rotation-based Doppler shifts well below the resolution limit, and surface magnetic field maps are not currently available \citep{Garraffo2016}. There is only a single measurement of the average magnetic field strength on Proxima \citep{Reiners2008}, which showed moderate magnetic flux of $450~\mathrm{G}<B_f<750$~G ($3\sigma$). The X-ray/UV intensity of the Proxima's emission anti-correlates with optical V-band brightness for both rotational and cyclical variations, possibly, and shows that all these variations are driven by magnetic activity. Optical intensities anti-correlates with the higher energy emission showing a minimum of magnetic activity (and minimum X-ray/UV emission) when the star is optically brightest (least spotty), unlike the relatively inactive Sun. The cycle 
amplitude of Prox Cen in X-rays is relatively small, with maximum and minimum X-ray luminosities $L_{\max\mathrm{X}}/L_{\min\mathrm{X}}$
roughly 1.5 versus 2--6 for the G and K stars \citep{Gudel2004}. 

The flare activity of Proxima Centauri is well known and was reported in a number of papers \citep{Thackeray1950,Walker1981,Haisch1983,Haisch1995,Gudel2004,Davenport2016}. According to recent MOST observations of flares on Proxima Centauri \citep{Davenport2016} flares with flux amplitudes of 0.5{\%} occur 63 times per day, while super flares with energies of $10^{33}$~erg occur 8 times per year. Comparing to other M5--M6 stars suggests Proxima was more active in its youth. A quiescent luminosity for Proxima Cen in the MOST bandpass is of $\log L_0 =28.69$~erg~s$^{-1}$.

\citet{Garraffo2016} constructed 3-D MHD models of the wind and magnetic field around Proxima Centauri using a surface magnetic field map for a star of the same spectral type and scaled to match the observed 600~G surface magnetic fieeld strength. They probed two different scalings of the magnetic field: field amplitude is equal 600~G and the mean magnetic field is 600~G so the maximum value is 1200~G). The wind speeds obtained by their model are 
not drastically different to the solar wind ones and consist up to 1300~km~s$^{-1}$ for the lower magnetic field case and up to 1600~km~s$^{-1}$ for the higher magnetic field case. The wind densities at Proxima b's orbital distance are 100 to 1000~cm$^{-3}$. 

The goal of this work is to estimate stellar wind properties, fluxes of galactic and stellar cosmic rays near Proxima Centauri~b accounting the stellar activity. We will use simple formulae with clear physical sense, which have been proposed in the beginning of space era. These formulae give answers with accuracy of factor 2--3 for the Sun and the Earth. Since the assumptions under which these formulae were derived do not dependent on star, we suppose that our results for Proxima Centauri would have the same accuracy. The reason for using such approach is that we don't know more about Proxima Centauri then the people knew about the Sun in the 50$^{\mathrm{th}}$ of the last century. Furthermore such simple modeling allows obtaining the full picture of processes in system without knowing details of the processes on the star and in the star's wind. Moreover such quality estimations by the order of magnitude can help to find relevant values of parameters and compare them with solar one. Such evaluations should allow 
simplifying subsequent equations and numerical simulations. 

\section{Stellar wind and astrosphere of Proxima Centauri}

According to \citet{Parker1958} we may estimate a sound speed depending on the coronal electron temperature as 
$u_{cr}=\sqrt{2kT_{e}/m_{p}}$, a radius of the critical point as
$r_{cr}={GM}/u_{cr}^{2}$ and stellar wind velocity as $V_{SW}\approx u_{cr}\ln(r_{b}/r_{cr})$.

Knowing a velocity of stellar wind we may also estimate its density. The rate of thermal loss is 
$$
Q=-\frac{8\pi}{7}R_{\ast }k(T_{c\ast})T_{c\ast },
$$ 
where $R_{\ast }$ is the stellar radius, $T_{c\ast}$ is the coronal 
temperature and $k(T_{c\ast })=6\times{10}^{-6}T_{c\ast}^{5/2}$~erg~cm$^{-1}$~s$^{-1}$~K$^{-1}$ is the coefficient of heat 
conductivity for a fully ionized gas. If all of the heat flux $Q$ went into an 
expanding spherical corona 
$$
Q\approx 4\pi r^{2}m_{p}\frac{NV}{2}(V_{SW}^{2}+V_{esc}^{2}), 
$$
where $N$ is the coronal density at distance $r$, $m_{p}$ is the proton mass, 
$V_{SW}$ is the constant stellar wind speed 
and $V_{\mathrm{esc}}=\sqrt{{2GM_{\ast}}/{R_{\ast}}}=568$~km~s$^{-1}$ is the escape velocity for Proxima Centauri.

This approach is summarized in \citet{Lang1980} and provides reasonable values for the Sun 
with accuracy of about one order of magnitude for the coronal temperature $10^{6}$~K.

 According to the X-ray observations the coronal temperature of Proxima is  $2.7\times{10}^{6}$~K (\citet{Johnstone2015}). Note that if $T_{e}\ge4.6\times{10}^{6}$~K a critical point would be below the surface of Proxima, i.~e. this is the maximal temperature for a quite corona. 
 
Knowing properties of the stellar wind we may estimate the radius of the Proxima astrosphere, that is $R_{\mathrm{AS}}=R_{b}(m_{p}nV^{2}/P_{\mathrm{ISM}})^{1/2}$, where $P_{\mathrm{ISM}}=0.17$~eV~cm$^{-3}$ is the energy density of local 
interstellar medium. The results of calculations of stellar wind parameters for different coronal temperatures are shown in Tab.~\label{tab0}. Even for the maximal possible coronal temperature of $4.6\times {10}^{6}$~K the stellar wind speed is 1200~km/s that is well below the wind speeds obtained by \citet{Garraffo2016}. It is not clear from the model of \citet{Garraffo2016}, how is it possible to change a position of the critical point accounting the coronal magnetic field?
\begin{table*}
	\centering
	\caption{Parameters of stellar wind.}
	\label{tab0}
	\begin{tabular*}{0.95\linewidth}{l @{\extracolsep{\fill}} cccccccc}
	\noalign{\smallskip}\hline\hline\noalign{\smallskip}
		{$T$, MK} & 
		{$Q$, erg} & 
		{$H/R_\ast$, $10^{-1}$}& 
		{$U_{\mathrm{cr}}$, $10^2$~km~s$^{-1}$} & 
		{$R_{\mathrm{cr}}/R_{\ast}$}& 
		{$V$, $10^{2}$~km~s$^{-1}$}& 
		{$N$, $10^3$~cm$^{-3}$}& {$R_{\mathrm{AS}}$ $10^2$~AU}\\
	\noalign{\smallskip}\hline\noalign{\smallskip}	
	2.7& 
	$7.026\times10^{27}$& 
	2.880& 
	2.158& 
	1.736& 
	8.065& 
	1.977& 
	4.261\\
	\textbf{4.6}& 
	$\mathbf{4.535\times10^{28}}$& 
	\textbf{4.907}& 
	\textbf{2.817}& 
	\textbf{1.019}& 
	\textbf{1.203}& 
	\textbf{4.320}& 
	\textbf{9.394} \\
	\noalign{\smallskip}\hline
	\end{tabular*}
\end{table*}

As Proxima seems to rotate in the habitable zone and in many papers conditions on it were compared with conditions near the Earth. However, for the solar wind we have natural parameter to normalize any solar system values---solar radii. If we normalize the distance from the Proxima to Proxima~b via the star radii we obtain that $R_{Pb}=0.049~\mbox{AU}=7.35\times10^{11}~\mbox{km}=72R_{P}$, so it is more reasonable to compare the 
conditions near Proxima~b with ones near the Mercury orbiting at 64 solar radii.

\section{Proxima Centauri modulation of galactic cosmic rays}

According to \citet{Parker19581} the GCR modulation by solar wind occurs inside 
the solar wind shell, which extends uniformly and with spherical symmetry, 
from a solar distance $r=r_{1}$ out to $r=r_{2}$. Assumed that solar wind has 
magnetic field $B$  and velocity $V_{SW}$, $l=2\times{10}^{11}$~cm.
The steady state cosmic ray density $j_{0}(\eta )$ inside the modulation 
shell is related to the galactic density $j_{\infty }(\eta )$ outside by
\[
j_{0}(\eta)=j_{\infty}(\eta)\exp\left\{-\frac{12V_{SW}\left(r_{1}-r_{2}\right)lZ^{2}e^{2}B^{2}(\eta +1)}{\pi 
^{2}m^{2}c^{5}\left[ \eta (\eta +2) \right]^{3/2}} \right\},
\]
where a particles have mass $m$, charge $Ze$, and kinetic energy $\eta mc^{2}$.

Reasonable values of $j_{0}(\eta )/j_{\infty}(\eta )$ 
(GCR modulation) near the Earth were obtained by \citet{Parker19581} assuming 
$B\approx2\times{10}^{-5}$~G, $r_{1}-r_{2}=4$~AU, 
$v=1000$~km/s. 

\begin{figure}[ht!]
	\centering
	\includegraphics[width=0.85\linewidth]{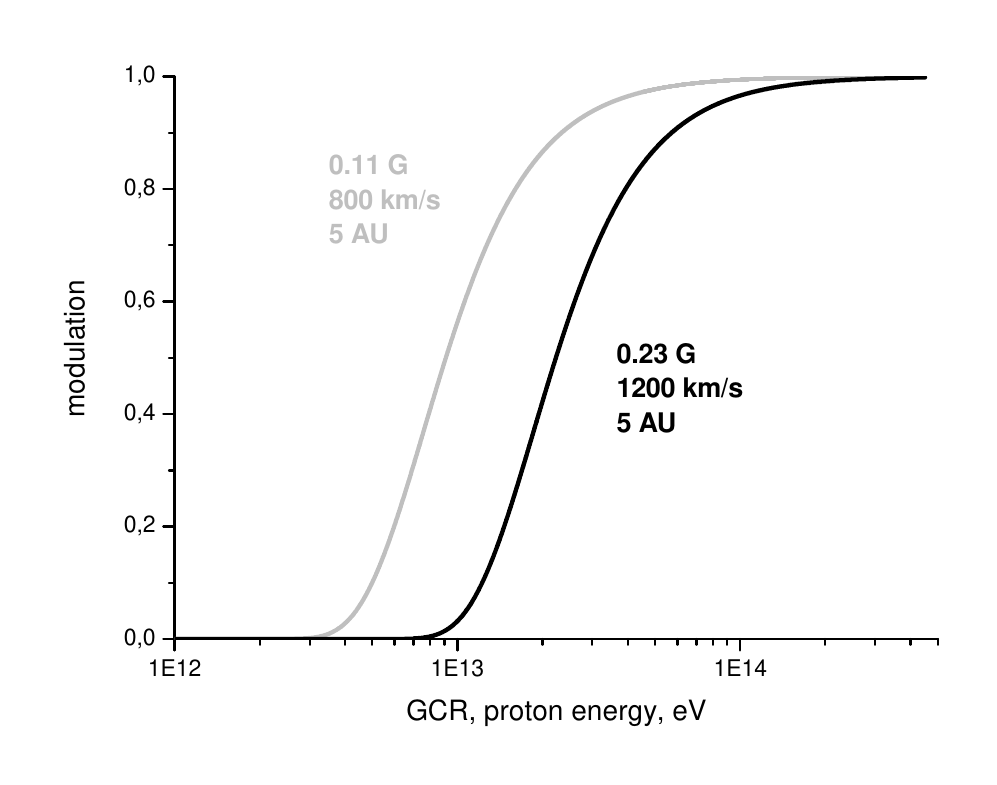}
	\caption{Calculated modulation ${j_{0}(\eta )}/j_{\infty}(\eta )$ for GCR near Proxima~b.}
	\label{fig1}
\end{figure}

In a case of Proxima~b $B\approx(1$--$2)\times{10}^{-1}$~G, $v=800$--1200~km/s. A radius of the Parker spiral for Proxima is 23--43~AU, within 
this range $B$ would not be constant, so for estimates we will take 
$r_{1}-r_{2}=5$~AU. Results of calculations are presented in Fig.~\ref{fig1}. We 
see that GCR protons with energies less than 1~TeV do not reach Proxima b, 
they are swept out by the stellar wind, the diffusion is not effective. It 
is clear that larger values of stellar wind velocity \citet{Garraffo2016} 
and shell should lead to stronger effects of GCR modulation.

\section{Stellar cosmic rays} 

The equipartition magnetic field strength is the minimum possible value for 
the field strength dynamic flare loops. Equating average magnetic field 
stellar surfaces to the field strength at which the magnetic field balances 
the thermal pressure of surrounding gas, $B^{2}/{8\pi} =nkT=Gm_{p}M_PH/R^{2}$, and in assumption that characteristic 
scale is equal to the mean free path $H=(n\sigma_{T})^{-1}$ ($\sigma 
_{T}$ is the Thompson cross-section), we obtain the estimation of the 
photospheric magnetic field strength 
$$
B_{0}=\frac{1}{R}\left(\frac{8\pi GmM}{\sigma_{T}}\right)^{1/2}.
$$

Maximal energy of Proxima protons accelerated in active region. Parameters 
of the typical active region are $L=\alpha R=\alpha0.145\times0.7\times10^{10}~\mbox{cm}=\alpha 10^{10}$~cm, $\alpha \lesssim1$; $B = \beta 
B_{0}\sim3150\beta$~G; $V =100$~km/c, and
\[
 E=\frac{1}{c}VB=\beta \times1\ \mbox{CGSE}, 
\]
\[
 U_{\max } =\alpha eER=3150\alpha \beta~\mbox{GeV}. 
\]
Maximal energy of Proxima CR is less than a minimal non-modulated energy of 
galactic CR. Spectra of stellar and galactic cosmic rays are not overlapped 
close to Proxima~b.

Flare energy we may express \citep{Balona2015,Struminsky2017} as 
$E_{fl}\alpha^{3}\beta^{2}=4.0\times10^{35}\alpha^{3}\beta^{2}$~erg. Therefore $10^{33}$~erg is a quite reasonable 
value of flare energy, according to the MOST observation \citep{Davenport2016} eight such flares per year occur, so their frequency is $f=3.2\times10^{-8}$~s$^{-1}$. Let 10{\%} of the flare energy were released into proton acceleration, then we may perform the Table~\ref{tab1}.

\begin{table}
	\centering
	\caption{Full number of protons in flare}
	\label{tab1}
	\begin{tabular*}{0.85\linewidth}{l @{\extracolsep{\fill}} c c}
	\noalign{\smallskip}\hline\hline\noalign{\smallskip}
	Energy, MeV & $N$, protons & $fN$, protons/s \\
	\noalign{\smallskip}\hline\noalign{\smallskip}
30 & 
$2\times10^{36}$& 
$6.4\times10^{28}$ \\
\hline
200 & 
$3\times10^{35}$ & 
$9.9\times10^{27}$ \\
	\noalign{\smallskip}\hline
	\end{tabular*}
\end{table}

Densities and fluxes of stellar cosmic rays in the modulation region (the 
first turn of the Parker spiral) are presented in Table~\ref{tab2}.

\begin{table*}
	\centering
	\caption{Densities and fluxes of stellar cosmic rays in the modulation region.}
	\label{tab2}
	\begin{tabular*}{0.95\linewidth}{l @{\extracolsep{\fill}} cccc}
	\noalign{\smallskip}\hline\hline\noalign{\smallskip}
	Energy, MeV &  \multicolumn{2}{c}{n, cm$^{-3}$} & \multicolumn{2}{c}{$nV/(2\pi)$, (cm$^{2}$~s~st$)^{-1}$} \\
	\noalign{\smallskip}\hline\noalign{\smallskip}
	&$V_{SW}=484$~km/s, 23~AU&$V=900$~km/s, 43~AU& $V_{SW}=484$~km/s, 23~AU & $V=900$~km/s, 43~AU\\
	\noalign{\smallskip}\hline\noalign{\smallskip}
30 & 
$1.2\times10^{-8}$& 
$1.8\times10^{-9}$& 
0.099& 
0.025 \\
200& 
$1.7\times10^{-9}$ & 
$2.7\times10^{-10}$ & 
0.013 & 
0.0038\\
	\noalign{\smallskip}\hline
	\end{tabular*}
\end{table*}

The main transport process of SCR is the convection. A characteristic time 
is $\tau = r/V = 4.2$~hours for $V= 484$~km/s (or 2.3~h for $V=900$~km/s).

If all accelerated protons will propagate within spatial angle $60\times60$~degrees, 
then fluence would be $F=9N/\pi r^{2}$ and the maximal intensity as 
$j_{\max}=F/(2\pi \tau) $ (Tab.~\ref{tab3}. Two different values of $j_{\max}$ in Table~\ref{tab3} correspond to different values of characteristic time.

\begin{table*}
	\centering
	\caption{Proton fluence for Proxima b.}
	\label{tab3}
	\begin{tabular*}{0.85\linewidth}{l @{\extracolsep{\fill}} cccc}
	\noalign{\smallskip}\hline\hline\noalign{\smallskip}
	Energy, MeV & $N$, protons & $F$, cm$^{-2}$&$j_{\max}$, (cm$^{2}$~s~st$)^{-1}$ & $j$, (cm$^{2}$~s~st$)^{-1}$ \\
	\noalign{\smallskip}\hline\noalign{\smallskip}
	30 & 
$2\times10^{36}$& 
$1.1\times10^{13}$& 
$1.2\times10^{8}$--$3.0\times10^{8}$& 
$1.6\times10^{11}\beta^{2}$ \\
200& 
$3\times10^{35}$& 
$1.6\times10^{12}$& 
$1.6\times10^{7}$--$4.4\times10^{7}$& 
$5.7\times10^{10}\beta^{\mathbf{2}}$ \\
	\noalign{\smallskip}\hline
	\end{tabular*}
\end{table*}

A 1/6 of the Proxima b year is 44 hours, i.~e. dynamics of SCR intensity 
would be determined by radial propagation of the stellar wind. The obtained 
fluencies are 2--3 orders more than the 30~MeV proton fluence of $8\times10^{10}$~cm$^{-2}$ 
estimated for the 775AD solar proton event. 

The obtained values of $j_{\max}$ we may compare 
with estimates according to formulae of \citet{Freier1963}, which would 
be $j=\beta^{2}B_{0}^{2} R^{4}v/(32\pi 
^{2}r^{4}E_{p})$ in our case, here $v$ is the proton velocity. We may get 
a coincidence between $j$ and $j_{\max}$ for reasonable values of $\beta $.

\section{Conclusions}

Cosmic rays are an important factor of space weather determining radiation conditions near planets so it is essential to know radiation conditions near extrasolar planets. 

We made estimates on parameters of stellar wind on the basis of the Parker model, possible fluxes and fluencies of galactic and stellar cosmic rays based on available data of the Proxima Centauri activity and its magnetic field. 

The simple models, which were derived for the Sun in 1950$^{th}$--1960$^{th}$, give the reasonable results for the star wind parameters and conditions on the orbit of Proxima b. For the first time and from the first principals with the help of available data the estimation of the radiation conditions near Proxima b was made. 

The obtained data showed that galactic cosmic rays will be absent near Proxima b up to energies till 1 TeV due to the modulation by the stellar wind. However stellar cosmic rays may be accelerated in stellar flares and swept out from the astrosphere by the wind. Flares at Proxima Centauri are able to maintain constant density of stellar cosmic rays in the astrosphere. Maximal proton intensities in extreme Proxima events should by 3--4 orders more than in solar events.

\section*{Acknowledgments}
{The work was partly supported by the Russian Foundation for Basic Research (grant 16-02-00328) and the Programm~1.7 P2 of the Russian Academy of Sciences.}

\bibliographystyle{ewass_ss4proc}

\begin{thebibliography}{}
	
	\bibitem[Anglada-Escude et al.(2016)]{Anglada2016} Anglada-Escude, G., Amado, P.~J., et al.\ 2016, Nature, 536, 437
	\bibitem[Atri(2017)]{Atri2017} Atri, D.\ 2017, MNRAS, 465, L34
	\bibitem[Atri et al.(2013)]{Atri2013} Atri,~D., Hariharan,~B., Grie{\ss}meier,~J.-M.\ 2013, Astrobiology, 13, 910 
	\bibitem[Balona(2015)]{Balona2015} Balona, L.~A.\ 2015, MNRAS, 447, 2714
	\bibitem[Cohen et al. (2012)]{Cohen2012}  Cohen, O.,  Drake, J. J., and  Kota J.\	2012, \apj, 760, 85 
	\bibitem[Davenport et al.(2016)]{Davenport2016} Davenport,~R. A.,. Kipping,~D.M, Sasselov,~D., et al.\
	2016, \apj, 829, L31
	\bibitem[Edwards \& McQueen(1971)]{Edwards1971} Edwards,~P.~J., \& McQueen,~M.\ 1971, Proc. 12th ICRC, 1, 323 
	\bibitem[Freier \& Webber(1963)]{Freier1963} Freier, P.~S., \& Webber, W.~B.\ 1963, JGR, 68, 1605 
	\bibitem[Garraffo et al.(2016)]{Garraffo2016} Garraffo, C., Drake, J.~J., Cohen, O.\ 2016, \apj, 833, L4
	\bibitem[Grie{\ss}meier et al.(2005)]{Griessmeier2005}	Grie{\ss}meier,~J.-M., Stadelmann,~A., Motschmann,~U., et al.\
	2005. Astrobiology, 5, 587 
	\bibitem[Grie{\ss}meier et al.(2015)]{Griessmeier2015} Grie{\ss}meier, J.-M., Tabataba-Vakili, F., Stadelmann, A., et al.\ 
	2015, \aap, 581, A44
	\bibitem[Grie{\ss}meier et al.(2016)]{Griessmeier2016} Grie{\ss}meier J.-M., Tabataba-Vakili, F., Stadelmann, A., et al.
	2016, \aap, 587, A159 
	\bibitem[G\"{u}del et al.(2004)]{Gudel2004} G\"{u}udel, M., Audard, M., Reale, F., et al.\
	2004, \aap, 416, 713 
	\bibitem[Haisch et al.(1983)]{Haisch1983} Haisch,~B.~M., Linsky,~J.~L., Bornmann,~P.~L., et al.\
	1983, \apj, 267, 280
	\bibitem[Haisch et al.(1995)]{Haisch1995} Haisch,~B.; Antunes,~A., Schmitt,~J.~H.~M.~M. 1995, Science, 268, 1327
	\bibitem[ Johnstone  \&  G\"{u}del (2015)]{Johnstone2015}
	 Johnstone, C., \& G\"{u}del, M.\ 2015, A\&A, 578, 129
	\bibitem[Kopysov \& Stozhkov(2005)]{Kopysov2005} Kopysov, Yu.~S., \& Stozhkov, Yu.~I.\ 2005, Proc. 29th ICRC, 3, 141
	\bibitem[Lang(1980)]{Lang1980} Lang, K.~R., 1980, Astrophysical Formulae, 2nd Edition (Berlin, Springer-Verlag ``Springer study edition''), 783	
	\bibitem[Lovell(1974)]{Lovell1974} Lovell, A.~C.~B.\ 1974, Phill. Trans. Roy. Soc., A 277, 489 
	\bibitem[Mullan(1979)]{Mullan1979} Mullan, D.~J.\ 1979, \apj, 234, 588
	\bibitem[Parker(1958)]{Parker1958} Parker, E.~N.\ 1958, \apj, 128, 664
	\bibitem[Parker(1958)]{Parker19581} Parker, E.~N.\ 1958, Phys. Rev., 110, 1445
	\bibitem[Reiners \& Basri(2008)]{Reiners2008} Reiners, A., Basri, G.\ 2008, \aap, 489, L45
	\bibitem[Scherer et al. (2002)]{Scherer2002} Scherer, K.,  Fichtner, H.,  Stawicki, O. \ 2002, J. of Atm.\& Solar-Ter. Phys., 64, 795
	\bibitem[Scherer et al.(2015)]{Scherer2015} Scherer, K., van der Schy,~A., Bomans, D.~J., et al.\ 
	2015, \aap, 576, A97 
	\bibitem[Stozhkov(2011)]{Stozhkov2011} Stozhkov, Yu. I.\ 2011, Bulletin of the Russian Academy of Sciences. Physics, 75, 323 
	\bibitem[Struminsky \& Sadovski(2017)]{Struminsky2017} Struminsky, A., \& Sadovski, A.\ 2017, APS Conference series, in print
	\bibitem[Tabataba-Vakili et al.(2016)]{Tabataba2016} Tabataba-Vakili, F., Grenfell, J.~L., Grie{\ss}meier, J.-M., Rauer, H.\ 2016, \aap, 585, A96 
	\bibitem[Thackeray(1950)]{Thackeray1950} Thackeray,~A.~D.\ 1950, MNAS of South Africa, 9, 9 
	\bibitem[Unsold(1957)]{Unsold1957} Uns\"{o}ld,~A.\ 1957, Proc. 4th IAU Symp., 238 
	\bibitem[Vidotto et al.(2014)]{Vidotto2014}  Vidotto, A.~A., Gregory, S.~G., Jardine, M., et al.\ 2014, MNRAS, 441, 2361
	\bibitem[Wargelin et al.(2017)]{Wargelin2017} Wargelin, B.~J., Saar, S.~H., Pojma?ski, G., et al.\ 2017, MNRAS, 464, 3281 
	\bibitem[Walker(1981)]{Walker1981} Walker, A. R.\ 1981, MNRAS, 195, 1029
	
	
\end{thebibliography}

\end{document}